\documentclass[12pt,preprint]{aastex62}
\usepackage{rotating}
\usepackage{epsfig}

\begin{document}


\title{A Radio Pinwheel Emanating from WR147}

\author{Luis F. Rodr{\'\i}guez}
\affiliation{Instituto de Radioastronom\'\i a y Astrof\'\i sica, 
Universidad Nacional Aut\'onoma de M\'exico,, Apdo. Postal 3-72 (Xangari), 58089 Morelia, Michoac\'an, M\'exico}

\affiliation{Mesoamerican Center for Theoretical Physics, Universidad
Aut\'onoma de Chiapas, Carretera Emiliano Zapata Km. 4,
Real del Bosque (Ter\'an), 29050 Tuxtla Guti\'errez, Chiapas,
M\'exico} 

\author{Jane Arthur}
\affiliation{Instituto de Radioastronom\'\i a y Astrof\'\i sica, 
Universidad Nacional Aut\'onoma de M\'exico,, Apdo. Postal 3-72 (Xangari), 58089 Morelia, Michoac\'an, M\'exico}

\author{Gabriela Montes}
\affiliation{Unaffiliated}

\author{Carlos Carrasco-Gonz\'alez}
\affiliation{Instituto de Radioastronom\'\i a y Astrof\'\i sica, 
Universidad Nacional Aut\'onoma de M\'exico,, Apdo. Postal 3-72 (Xangari), 58089 Morelia, Michoac\'an, M\'exico}

\author{Jes\'us A. Toal\'a}
\affiliation{Instituto de Radioastronom\'\i a y Astrof\'\i sica, 
Universidad Nacional Aut\'onoma de M\'exico,, Apdo. Postal 3-72 (Xangari), 58089 Morelia, Michoac\'an, M\'exico}

\email{l.rodriguez@irya.unam.mx}
 
\begin{abstract}
Wolf-Rayet (WR) stars are evolved massive stars, presumably on their way to becoming supernova. They are characterized
by high luminosities and fast and dense stellar winds. We have detected signs of a radio continuum pinwheel associated
with WR147, a nitrogen-rich WR star with spectral subtype WN8. These structures are known around a handful of late-type carbon-rich WR stars with massive companions were the dust formed at the zone where the two winds collide produces
a plume of dense gas and dust that is carried out with the WR wind. As the binary system rotates, an Archimedean spiral detectable
in the infrared is formed. The resulting pinwheel contains information on wind speeds, wind-momentum ratio and orbital parameters.
However, WR147 is a WN star and the formation of dust is unlikely so a different emission mechanism must be at work.
Our analysis of the data suggests that in this case the emission is dominantly
of non-thermal nature (synchrotron), although we cannot exclude that some clumps could be brighter in
free-free emission. It is possible that the pinwheels associated with WN stars will be detectable only as non-thermal emitters at
radio wavelengths. From the characteristics of the pinwheel we estimate a period of 1.7 years for the binary system (the WN8 star and a companion yet undetected directly) responsible for the pinwheel.


\end{abstract}  


 
\section{Introduction}

The Wolf-Rayet (WR) stars, discovered in 1867 by Charles Wolf and Georges Rayet, are believed to represent the final stages in the evolution of very massive stars. These stars will end their lives as supernovae. WR stars exist in two main types: those with 
strong lines of nitrogen and helium (WN subtype) and those with strong carbon and helium lines (WC subtype). A third WO subtype with oxygen and carbon lines has been proposed (e.g. Crowther 2007; Tramper et al. 2015).

Over the years it has become clear that late-type carbon-rich WR stars with massive binary companions can produce spiral pinwheel structures detectable as infrared sources. The accepted model for these structures is that dust forms at the high density
interface between the colliding stellar winds. A plume of dust-rich material is then injected into the surrounding medium and,
as the orbital motion wraps the plume, the pinwheel is formed. 
The first infrared pinwheel associated with a WR star was the case of WR104 (Tuthill et al. 1999). Over the years
a few additional cases have been reported: WR 98a (Monnier et al. 1999),  WR 112 (Marchenko et al. 2002),
WR140 (Monnier et al. 2002), Apep (Callingham et al. 2019) and possibly WR118 (Millour et al. 2009). All these cases involve a WC star with an OB companion (in the case
of Apep the binary is constituted by a WC star and a WN star, Callingham et al. 2020).This is not unexpected, since the WC winds are capable of nucleating graphite (a crystalline form of carbon and key component of interstellar dust) in the shock regions. In contrast, no WN star has been found to be associated with an infrared pinwheel structure.

It is known that relativistic electrons can be produced in the wind collision zone of massive binaries
(e.g. White \& Becker 1983; Pittard 2010,
De Becker \& Raucq 2013). Following the reasoning for the dust, one could expect to
detect synchrotron-emitting pinwheels in the case of WN stars. In this paper we present high angular resolution, high sensitivity
radio observations of the WN star WR147 that indeed reveal a radio pinwheel of non-thermal nature.

\section{Observations}

The observations were part of our
VLA project 13B-258, made with the Karl G. Jansky Very Large Array (VLA) of NRAO\footnote{The National 
Radio Astronomy Observatory is a facility of the National Science Foundation operated
under cooperative agreement by Associated Universities, Inc.} in its highest angular resolution A-configuration
during 2014 February 24. 
The flux and bandpass calibrator was J0137+331 and
the phase calibrator was J2007+4029.
The digital correlator of the VLA was configured in 48 spectral windows of 128 MHz width, covering the range of
12 to 18 GHz. Each spectral window is divided 
in 64 channels with individual spectral resolution of 2 MHz. 
The data were calibrated in the standard manner using the CASA (Common Astronomy Software Applications;  McMullin et al. 2007) package of NRAO and
the pipeline provided for VLA\footnote{https://science.nrao.edu/facilities/vla/data-processing/pipeline} observations. In addition, the data were self-calibrated in phase and amplitude.
Maps were made using a robust weighting (Briggs 1995) of -0.5, searching to optimize the compromise between sensitivity and angular resolution. 

\section{Interpretation}

In Figure 1 we show a color image of the WR147 region made to emphasize the faint emission that
traces the radio pinwheel. In Figure 2 we show the same data in the form of a contour image.
Two well-known sources dominate the region: The first one, to the south, traces the thermal (free-free) emission of the wind from WR147.
The northern one, bow-shaped, traces the non-thermal (synchrotron) emission produced in the interaction zone of the
winds of WR147 with its B0.5V (Williams et al. 1997) companion. The spectral type of the northern companion
is uncertain, with Niemela et al. (1998) favoring an O8-O9 V-III spectral type. We will refer to it as the OB star.
This northern emission appears slightly to the south of the 
OB star since it traces the wind interaction zone and not the star itself (Abbott et al. 1986; Moran et al. 1989; Churchwell et al. 1992; Contreras et al. 1997; Williams et al. 1997; Skinner et al. 1999; Watson et al. 2002). It is known that these two sources are physically related because of the morphology of the bow shock and their common proper motions (Dzib \& Rodriguez 2009).

The new component is the spiral structure that appears to emanate from WR147. With respect to WR147, the spiral
starts at a position angle of $\sim-60^\circ$ and turns counterclockwise. When it reaches a position angle (PA) of $\sim$180$^\circ$, 
the emission
becomes faint and is detected only as a sequence of 4-$\sigma$ blobs. Interestingly, when the spiral reaches a PA of
$\sim$340$^\circ$, it seems to interact with the non-thermal bow shock located close to the northern OB companion
and becomes relatively bright again. However, it is unclear if this enhancement is due to a physical interaction between both
structures or to the superposition of their emissions.
Finally, after a position angle of $\sim$380$^\circ$,
the pinwheel is no longer detectable. Position angles larger than 360$^\circ$ refer to points in the second crossing of the spiral with
respect to the north.

Using the multi-frequency continuum imaging capability of CASA (Rau \& Cornwell 2011)
we can measure directly the spectral indices  of WR147 and of the northern bow shock. These spectral indices are
given in Table 1. The spectral index of the wind of WR147, 0.55$\pm$0.04, is typical of
an ionized thermal (free-free) wind (e.g. Panagia \& Felli 1975). In contrast, the spectral index of the northern bow shock,
$-$1.0$\pm$0.1, is characteristic of optically-thin synchrotron. Determining the spectral index of the pinwheel is not
as straightforward, given its faintness. We can measure the spectral indices directly with significant signal-to-noise
ratio only in the segments spanning from
$-$50$^\circ$ to $0^\circ$ and 340$^\circ$ to $350^\circ$, obtaining values of $-$0.8$\pm$0.2 and $-$1.0$\pm$0.4, respectively.
To estimate the spectral index of the pinwheel as a whole we used our data at 22 GHz, also from project 13B-258,
to make an image with the same angular resolution as the 15 GHz data presented here and determined the flux densities inside
a polygon enclosing all the detectable pinwheel emission at 15 GHz. The pinwheel structure is marginal at 22 GHz, but
a determination of its total flux density is possible. We obtain flux densities of 3.5$\pm$0.1 mJy at 15 GHz
and of 2.3$\pm$0.1 mJy at 22 GHz. This implies a spectral index of $-$1.1$\pm$0.1, consistent with the value measured in the
northern bow shock and in the brightest parts of the pinwheel.

The models for the non-thermal emission from a colliding-wind binary of Pittard et al. (2020)
predict "cooling" of the relativistic electrons as they
flow away from the shock. This cooling should be manifest in changes in the brightness and
spectral index with distance from the shock, but 
our measurements are not accurate enough to compare with the models. A rough calculation following
Condon \& Ransom (2016) and assuming energy equipartition and optically-thin synchrotron emission
gives a minimum magnetic field of $\sim$5 mGauss for the segment spanning from
$-$50$^\circ$ to $0^\circ$. A decay lifetime of order $10^3$ years is also derived for the relativistic electrons.
Under these assumptions the pinwheel can remain detectable over a relatively long distance from the star. 
However, this lifetime estimate assumes only losses by synchrotron radiation and is
an upper limit since other mechanisms (i. e. inverse Compton scattering of the stellar photons or expansion of the emitting
region) could also make the radio emission decrease.

\clearpage

\begin{figure}
\centering
\vspace{-0.0cm}
\includegraphics[angle=-90,scale=0.5]{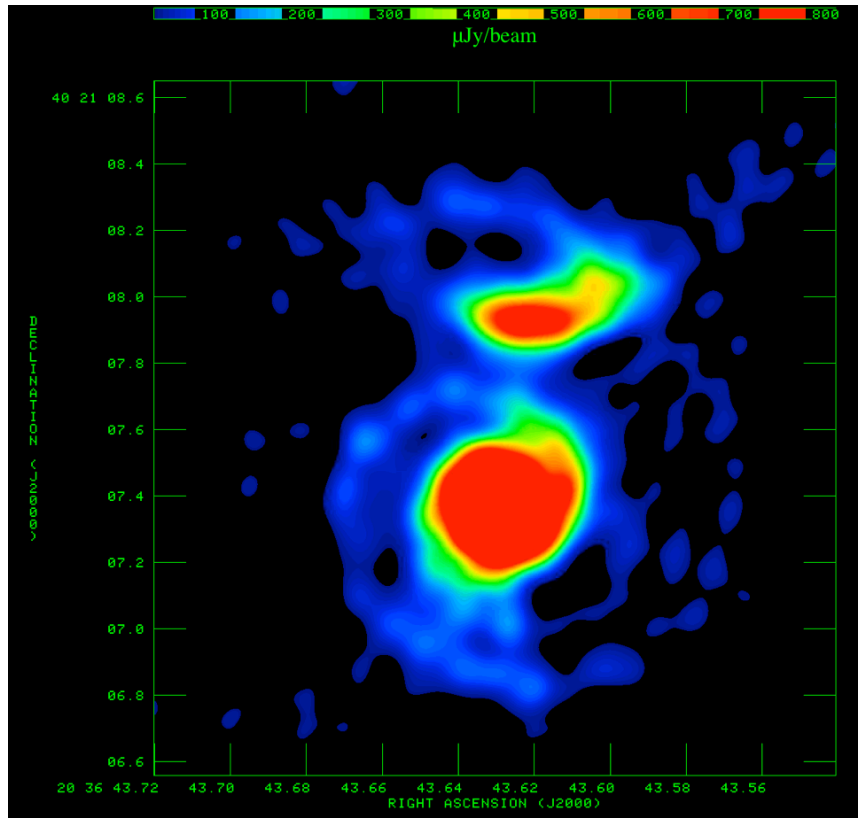}
\vspace{1.0cm}
\caption{\small False-color image of WR147 and its surroundings at 15 GHz. The color range is shown in the bar at the top of
the figure and goes from 25 to 
800 $\mu$Jy beam$^{-1}$, with all flux densities above the latter value appearing saturated in red. This selection of color parameters has the purpose of emphasizing the faint spiral feature in the region. 
This same data is shown as a contour image in Figure 2
}

\label{fig1}
\end{figure}

\clearpage

\begin{figure}
\centering
\vspace{-2.8cm}
\includegraphics[angle=0,scale=0.7]{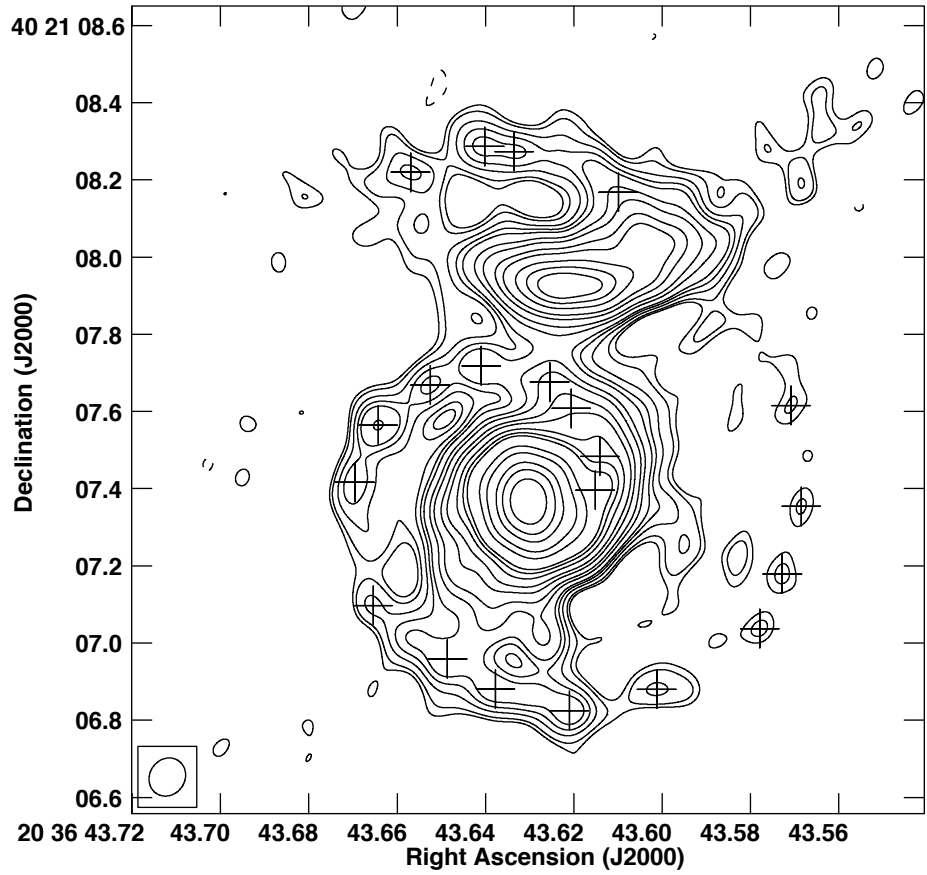}
\vspace{-2.8cm}
\caption{\small Contour image of WR147 and its surroundings at 15 GHz. The beam ($0\rlap.{''}10 \times 0\rlap.{''}09$; PA = -34$\circ$) is shown in the bottom left corner.
The contours are -3,  3, 4, 6, 8, 10, 15, 20, 30, 40, 60, 80, 100, 200, 300, 500, 700, and 1000 times 8 $\mu$Jy beam$^{-1}$, the
rms of the image. The crosses mark the positions of the local maxima used to describe the pinwheel.
}

\label{fig2}
\end{figure}

\clearpage

\clearpage

\begin{figure}
\centering
\vspace{-1.8cm}
\includegraphics[angle=0,scale=0.45]{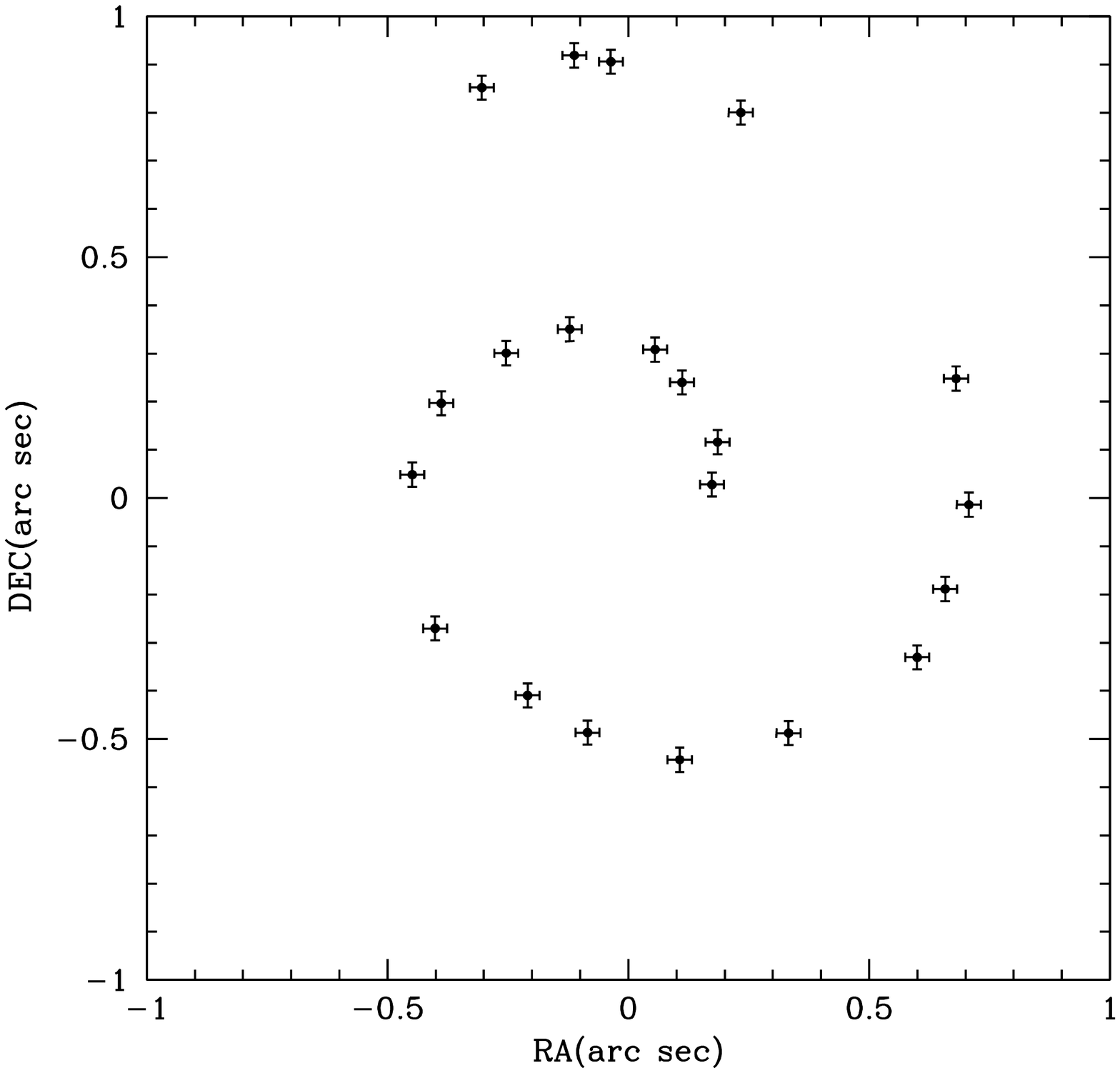}
\vskip-3.7cm
\includegraphics[angle=0,scale=0.45]{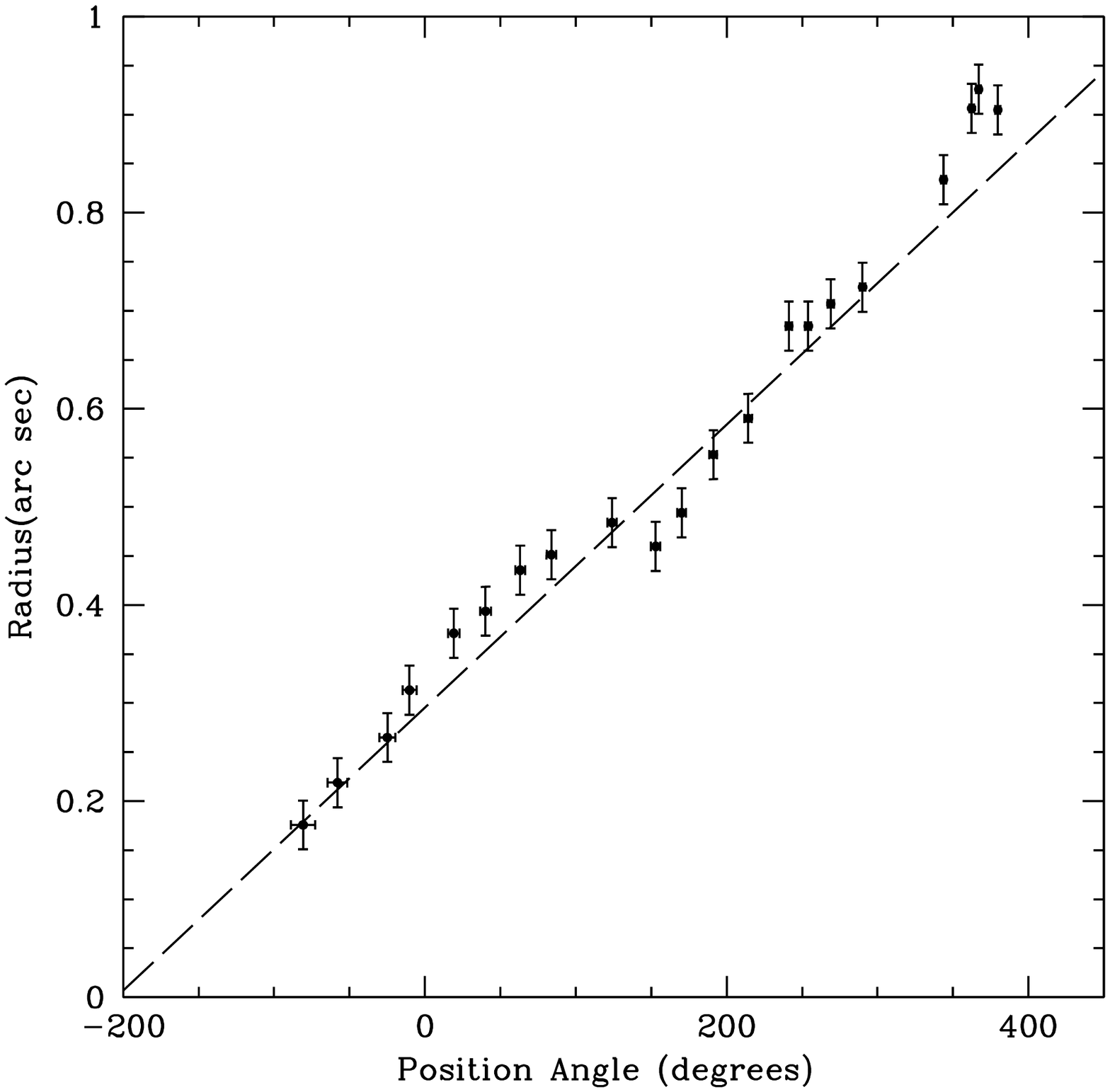}
\vskip-1.3cm
\caption{\small (Top) Positions of local maxima across the radio pinwheel with respect to the position of
the peak emission of WR147, RA(2000) = $20^h 36^m 43\rlap.^s630$; DEC(2000) = $40^\circ 21' 07\rlap.{''}37$.
(Bottom) Radius of the pinwheel as a function of position angle. The dashed line is the least-squares fit described in the text.
}

\label{fig5}
\end{figure}

\clearpage

\begin{deluxetable}{lcc}
\tablecaption{Flux Densities and Spectral Indices of the Sources at 15 GHz}
\tablehead{
\colhead{} & \colhead{Flux} & \colhead{Spectral} \\
\colhead{Source} & \colhead{Density (mJy)} & \colhead{Index}
 }
\startdata
WR147 &  35.3$\pm$0.1 & +0.55$\pm$0.04\\
Bow Shock &  3.6$\pm$0.1 & $-$1.0$\pm$0.1\\
Pinwheel &  3.5$\pm$0.1 & $-$1.1$\pm$0.1$^a$\\
\enddata
\tablenotetext{a}{Determined from a combination of 15 and 22 GHz data.}
\end{deluxetable}

To characterize the pinwheel structure we determined the local maxima along the spiral, indicated with the crosses
in Figure 2. In Figure 3 we present an $(x,y)$  plot with the positions of these maxima.
The spiral appears to be of the Archimedean kind (radius proportional to angle) and in Figure 3 we also show 
a radius versus position angle plot derived from the positions using the equations 

$$r = (x^2 + y^2)^{1/2};   ~~~~~~~~   PA = -tan^{-1}(x/y). $$

As can be seen, the radius versus position angle plot can be approximately described by a straight line, supporting the
Archimedean interpretation. A least-squares fit to these data gives:

$$\left[r \over '' \right] = 0.30 \pm 0.01 + 0.53  \pm 0.01 \left[PA \over 360^\circ \right]. $$

\section{Determination of the period of the binary producing the pinwheel}

WR147 is located at a photometric distance of 650 pc (Morris et al. 2000). 
It is interesting to note that the Gaia parallax of WR 147 is negative (Gaia Collaboration et al. 2018) presumably
because of the multiple nature of this object. From the fit derived above, the angular separation between
consecutive turns of the pinwheel is $0 \rlap.{''}53  \pm 0 \rlap.{''}01$, that corresponds to a physical scale of
340$\pm$7 au. Assuming that the pinwheel expands at  950 km s$^{-1}$, the velocity of the wind of WR147
(Morris et al. 2000), we obtain a period of $\sim$1.7 years.

This result, combined with information available in the literature, suggests that WR147 is at least a triple system.
We first have the WN8 star itself, the dominant source in the system. About $0 \rlap.{''}64$ to its north we have the OB star
whose wind, in interaction with that of the WN8 star produces the well-known non-thermal bow shock. Assuming a total mass
of 40 M$_ \odot$ for the pair and using Kepler's third law, we roughly estimate a period of order 1,300 years for this system.
We then have the star responsible, with the WN8 star, of producing the pinwheel, forming a system with an estimated period of 1.7 years.
Assuming a total mass
of 30 M$_ \odot$ for this pair and using again Kepler's third law, we estimate a separation of $\sim$4 au ($0\rlap.{''}006$) 
for the binary that produces the pinwheel.

Using {\it Chandra} observations, Zhekov  \& Park (2010a) determined that WR147 is also a double X-ray source,
with the X-ray components corresponding to the main radio ones. Although single WN2--6 are X-ray emitters, latter WN7--9 
types are difficult to detect (Skinner et al. 2010, 2012; Toala et al. 2018), prompting Zhekov \& Park (2010a) to propose the presence of a companion.
Zhekov  \& Park (2010b) detected a periodicity of 15 days in the X-ray emission from the
WN8 star and proposed that this implies the presence of a very close star to WR147 in the system. If confirmed, this would be a fourth
star in the system. However, it is known that massive stars can present variability in their X-ray emission due to other
phenomena (Oskinova 2016)  and the reality of this fourth star is uncertain.

The finding that radio pinwheels can be detected in association with WN stars opens the possibility of obtaining information
on the orbital periods, wind speeds and wind-momentum ratios in binaries containing this subclass of stars. As in the case of infrared pinwheels (Callingham et al. 2019), the detection
of unusual patterns in the radio pinwheels may point to progenitors of long-duration gamma ray bursts.

\acknowledgements
We thank an anonymous referee for a careful revision of our paper that improved its clarity.
LFR acknowledges the financial support of
PAPIIT-UNAM and of CONACyT (M\'exico).
This research has made use of the SIMBAD database,
operated at CDS, Strasbourg, France.

\facility{VLA}

\software{AIPS van Moorsel et al. (1996); CASA McMullin et al.(2007)}

\end{document}